\documentclass[11pt]{article}
\usepackage{enumitem}
\usepackage{natbib}
\usepackage{amsmath,amsthm,amssymb}
\usepackage{mathrsfs}
\usepackage{booktabs}
\usepackage{graphicx,psfrag,epsf}
\usepackage{epstopdf}
\usepackage{epsfig}
\usepackage{subcaption}
\usepackage{bbm}	
\usepackage{color}
\usepackage{mathtools} 
\usepackage{authblk}
\usepackage{tikz}
\usetikzlibrary{arrows}

\RequirePackage[hyperindex,breaklinks,colorlinks,citecolor=blue,urlcolor=black]{hyperref} 

\setitemize{itemsep=-1mm}
\setenumerate{itemsep=-1mm}

\theoremstyle{plain}

\newtheorem{theorem}{Theorem}
\newtheorem{assumption}{Assumption}

\theoremstyle{definition} 
 
\newtheorem{example}{Example}

\renewcommand{\eqref}[1]{(\ref{#1})}

\newcommand{\mtt}[1]{\texttt{\textsc{#1}}}
\newcommand\indep{\protect\mathpalette{\protect\independenT}{\perp}}
\newcommand{\aipwhat}{\hat{\tau}(\hat{\alpha}, \hat{\beta}_0, \hat{\beta}_1)}
\newcommand{\aipwst}{\hat{\tau}(\alpha^*, \beta_0^*, \beta_1^*)}

\newcommand{\RN}[1]{%
  \textup{\uppercase\expandafter{\romannumeral#1}}%
}
\def\independenT#1#2{\mathrel{\rlap{$#1#2$}\mkern2mu{#1#2}}}

\newcommand{\blind}{0}

\makeatletter
\newcommand{\mylabel}[2]{#2\def\@currentlabel{#2}\label{#1}}
\makeatother

\begin{document}

\def\spacingset#1{\renewcommand{\baselinestretch}%
{#1}\small\normalsize} \spacingset{1}


\if0\blind
{
      \title{\bf Variable Selection for Doubly Robust Causal Inference
  }
\author[1]{Eunah Cho\thanks{ Corresponding author: \texttt{eunah.cho@lgdisplay.com}}}
\author[2]{Shu Yang}
\affil[1]{AI/Big Data Analysis Team, LG Display}
\affil[2]{Department of Statistics, NC State University}
\date{}
  \maketitle 
} \fi

\if1\blind
{
  \bigskip
  \bigskip
  \bigskip
  \begin{center}
    {\LARGE\bf Title}
\end{center}
  \medskip
} \fi

\begin{abstract}
Confounding control is crucial and yet challenging for causal inference
based on observational studies. Under the typical unconfoundness assumption,
augmented inverse probability weighting (AIPW) has been popular for
estimating the average causal effect (ACE) due to its double robustness
in the sense it relies on either the propensity score model or the
outcome mean model to be correctly specified. To ensure the key assumption
holds, the effort is often made to collect a sufficiently rich set of pretreatment
variables, rendering variable selection imperative. It is well known
that variable selection for the propensity score targeted for accurate prediction
may produce a variable ACE estimator by including the instrument variables.
Thus, many recent works recommend selecting all outcome predictors
for both confounding control and efficient estimation. This article shows that the AIPW estimator with variable selection targeted for
efficient estimation may lose the desirable double robustness property.
Instead, we propose controlling the propensity score model for any
covariate that is a predictor of either the treatment or the outcome
or both, which preserves the double robustness of the AIPW estimator.
Using this principle, we propose a two-stage procedure with penalization
for variable selection and the AIPW estimator for estimation. We show the proposed
procedure benefits from the desirable double robustness property. We evaluate
the finite-sample performance of the AIPW estimator with various variable
selection criteria through simulation and an application.
\end{abstract}

\noindent {\it Keywords: Confounding; Covariate selection; Double robustness}

\spacingset{1.0}

\section{Introduction}
\subsection{Ignorability and the need for variable selection }
Unlike experimental studies, treatment assignments in observational
studies are not random. As a result, distributions of the covariates differ between treatment arms, and direct comparisons between the treatment groups may be biased. Most causal inference methods rely on the ignorability assumption (also referred to as no unmeasured confounders), which indicates the treatment assignment can be ignored when conditioning the observed covariates. Under the ignorability assumption, researchers have proposed various methods to estimate the average causal effect (ACE), including regression imputation estimator, matching estimator, inverse propensity score weighted (IPW) estimator, augmented IPW (AIPW) estimator \citep[e.g.,][]{imbens2015causal}. Among them, the AIPW estimator has been popular because it is locally efficient and doubly robust in the sense that its consistency relies on either the correctly specified propensity score (PS) model or the outcome mean (OM) model, but not necessarily both \citep{robins1994estimation, rotnitzky1998semiparametric, scharfstein1999adjusting, lunceford2004stratification,bang2005doubly}. 

In the past, it was desirable to include all possible pretreatment variables to avoid the risk of excluding related variables and satisfy the ignorability assumption \citep{lunceford2004stratification,shortreed2017outcome}. With the advances in technology, a rich set of pretreatment covariates can be collected. In these high-dimensional settings, including all variables can be computationally unstable, burdensome, or sometimes impossible. Thus, variable selection is indispensable for handling high-dimensional covariates.

\subsection{Existing variable selection strategies }

Typically, there are four main types of pretreatment variables: 1) instrumental variables, 2) confounders, 3) precision variables, and 4) spurious variables. We refer to variables that are only predictors of the treatment but not the outcome as the instrumental variables, variables that are predictors of both the treatment and outcome as the confounders, variables that are only predictors of the outcome but not the treatment as the precision variables, and others as the spurious variables. 

\citet{lunceford2004stratification} show that containing the precision variables in the PS model helps reduce standard errors while maintaining consistency. Following this result, many researchers suggest outcome predictor approaches, which include the precision variables
and the confounders for the ACE estimation, since other variables, including the instrumental
variables, may inflate the variance of the ACE estimators or introduce
bias to the estimator \citep{brookhart2006variable,patrick2011implications}.

 \citet{shortreed2017outcome} suggest the outcome-adaptive lasso and provide simulation studies
showing that including precision variables in the PS model increases efficiency. \citet{ertefaie2018variable} propose a penalized objective
function that simultaneously considers the outcome and treatment assignment
models for variable selection. \citet{tang2020outcome} propose the
causal ball screening that targets confounders and other outcome predictors
as an adjustment set for the PS. \citet{henckel2019graphical}
provide a pruning procedure determining the optimal adjustment set.
They show the adjusted least-squares treatment effect estimator based
on the identified set has the smallest asymptotic variance among consistent
adjusted least square estimators. \citet{rotnitzky2019efficient}
demonstrate that Henkel's results can also be extended to non-parametric estimators.
However, it is challenging
to distinguish  confounders from  instrumental variables, and
hence the outcome predictor approach may exclude the true confounders
from subsequent estimation. As a result, the omission of important
confounders may lead to bias. To avoid such bias, \citet{vanderweele2011new} suggest
adjusting for any covariates that are causes of either the treatment
or outcome because those variables constitute a sufficient set
adjusting for confounding. 
\citet{belloni2014inference} suggest a post-double-selection method where they consider  the union of the covariates considered important in two equations from a partially linear model and estimate the ACE using linear square regression.
\citet{wilson2014confounder} estimate the standard Bayesian regression model and then the posterior distribution using a confounder-specific loss function. They target the set of all confounders and the outcome-related covariates, but instrumental variables are included to avoid omitting confounder variables. Although the idea of using the union of the selected variables has already been proposed to avoid bias due to the exclusion of confounders, we provide another reason for using the union approach in terms of maintaining the double-robustness of the AIPW estimator. 

\subsection{Contribution and outline}
This article shows that the AIPW estimator with variable selection targeted for efficient estimation (referred to as the outcome predictor approach) may lose the desirable double-robustness property. Generally, the outcome predictor approach is shown to be efficient, provided the postulated models are known or correctly specified. However, if the PS model
is correctly specified, and the OM model is misspecified, the estimation of the PS model restricted to the outcome predictors may not be consistent; see Example \ref{example 1}. Thus, although the PS working model is correctly specified, the PS model based on the wrong set from the OM model is not consistent for the true PS model, and thus the AIPW estimator becomes not consistent. 
 Given the above reasons, considering the selected instrumental variables in subsequent estimation aids in protecting the AIPW estimator's double-robustness property.   

Using this principle, we propose a two-stage procedure with penalization
for variable selection and the AIPW estimator for estimation. In the first stage,
we select a set of variables considered important predictors of either
the treatment or outcome using penalized estimating equations.
In this paper, we used the smoothly clipped absolute deviation (SCAD) proposed by \cite{fan2001variable}, but other penalized methods also can be
applied. After variable selection, we employ the AIPW estimator to
estimate the ACE with the nuisance models refitted based on the selected
variables. We show the proposed procedure benefits from the desirable statistical
properties, including selection consistency and double robustness. 

The rest of the paper is organized as follows. Section \ref{2 setup}
presents the basic setup. Section \ref{3 method} illustrates the
wishlist consisting of various variable selection criteria. Section
\ref{4 asymptotic} presents the asymptotic properties of our procedure.
In Section \ref{5 simulation}, we compare our approach to common
variable selection strategies for confounding control or efficient
estimation. The simulation suggests that the AIPW estimator is still
doubly robust with our variable selection procedure but is not with
other selection strategies. In Section \ref{6 application}, we apply our procedure to an application, 
maternal smoking on birth weight data. We conclude the paper with a discussion
in Section \ref{7 sec:Concluding-remarks}.

\section{Basic Setup } \label{2 setup}

\subsection{Potential outcomes framework}

Following \cite{neyman1923applications} and \cite{rubin1974estimating},
we adopt the potential outcomes framework. Denote $X$ to be a vector
of $p$-dimensional pretreatment covariates. Suppose that the treatment
is a binary variable $A\in\{0,1\}$, with $0$ and $1$ being labels
for control and active treatments, respectively. Under the common
Stable Unit Treatment Value assumption \citep{rubin1980randomization},
for each level of the treatment $a$, we assume that there exists
a potential outcome $Y(a)$, representing the outcome had the unit,
possibly contrary to the fact, been given the treatment $a$. We make
the consistency assumption that links the observed outcome with the
potential outcomes, i.e., the observed outcome $Y$ is the potential
outcome under the treatment regime actually following $Y(A)$. We focus
on estimating the ACE, $\tau=E\{Y(1)-Y(0)\}$. The ACE is the target
causal estimand in many scientific applications, generating important
policy implications. Our methodology also applies to a broader class
of causal estimands in \cite{li2018balancing}.

The fundamental problem in estimating the ACE is that one may observe
at most one of $Y(0)$ and $Y(1)$ for each unit. Throughout, we make
the ignorability assumption widely used in the causal inference
literature. 
\begin{assumption}[Ignorability] \label{Ignorability} $\{Y(0),Y(1)\}\indep A\mid X$.\end{assumption} 
\begin{assumption}[Overlap] \label{Overlap} There exist constants
$c_{1}$ and $c_{2}$ such that $0<c_{1}\leq e(X)\leq c_{2}<1$ almost
surely, where $e(X)=P(A=1|X)$ is the PS. \end{assumption} 
Assumption \ref{Ignorability} holds when all confounders are identified
and measured. That is, this assumption requires $X$ to include
all factors related to both treatment and outcomes. Assumption \ref{Ignorability}
indicates that treatment assignment is independent of the potential
outcomes given $X$. This assumption is automatically guaranteed in
the experimental study because the treatment is assigned to each unit
at random. In observational studies, researchers often collect a rich
set of pretreatment covariates to make this assumption plausible,
leading to possibly high-dimensional $X$. 

Assumption \ref{Overlap} holds when there exists a sufficient overlap
between the covariate distributions of the treatment and control group.
This means that the distributions of the treatment and control groups
need to be similar to each other. When Assumption \ref{Overlap}
is not satisfied at a specific value of $X$, the unit at the
value would be only treated or controlled. This leads to the extrapolation
of one of two potential outcomes at that value and makes the inference
about the ACE inappropriate.

\subsection{Doubly robust estimator of the ACE}

It is well known that under Assumption \ref{Ignorability} and Assumption
\ref{Overlap}, the ACE can be identifiable and estimated through the
outcome regression or the augmented/inverse probability weighting
(AIPW/IPW) estimator. See \cite{imbens2004nonparametric} and \cite{rosenbaum2002overt}
for surveys of these estimators.

Define $\mu_{a}(X)=E\{Y(a)|X\}$ for $a=0,1$. Then, under Assumption
\ref{Ignorability}, $\mu_{a}(X)=E(Y|A=a,X)$. In practice, the outcome
distribution and the PS are often unknown and therefore
have to be modeled and estimated. 

\begin{assumption}[Outcome mean model] \label{outcome} The parametric
model $\mu_{a}(X)=E(Y|A=a,X)$ is a correct specification for $\mu_{a}$(X),
for a=0,1; i.e., $\mu_{a}(X)=\mu_{a}(X;\beta_{a0})$, where $\beta_{a0}$
is the true OM model parameter for $a=0,1$. \end{assumption}

\begin{assumption}[Propensity score model] \label{PS} The parametric
model $e(X;\alpha_{0})$ is a correct specification for $e(X)$; i.e.,
$e(X)=e(X;\alpha_{0})$, where $\alpha_{0}$ is the true model parameter.
\end{assumption} 

Under Assumption \ref{outcome}, let $\hat{\beta}_{a}$ be a consistent
estimator of $\beta_{a0}$. Under Assumption \ref{PS}, let $\hat{\alpha}$
be a consistent estimator of $\alpha_{0}$. The\textit{ Augmented
Inverse Propensity score Weighting (AIPW) estimator} is
\begin{multline*}
\widehat{\tau}_{n,\text{AIPW}}=\frac{1}{n}\sum_{i=1}^{n}\Bigg[\frac{A_{i}Y_{i}}{e(X_{i};\hat{\alpha})}+\bigg\{1-\frac{A_{i}}{e(X_{i};\hat{\alpha})}\bigg\}\mu_{1}(X_{i};\hat{\beta}_1)\\
-\frac{(1-A_{i})Y_{i}}{1-e(X_{i};\hat{\alpha})}-\bigg\{1-\frac{(1-A_{i})}{1-e(X_{i};\hat{\alpha})}\bigg\}\mu_{0}(X_{i};\hat{\beta}_0)\Bigg].
\end{multline*}
The AIPW estimator is doubly robust in the sense that it is consistent
if either Assumption \ref{outcome} or \ref{PS} holds and locally
efficient if both assumptions hold \citep{rotnitzky2014double}.

Without loss of generality, we assume all covariates have a mean of
zero and a common standard deviation so that we apply the penalty
equally to all covariates. Define $\alpha^{*}=\text{argmin}_{\alpha\in\mathbb{R}^{p}}E[\{A-e(X^{T}\alpha)\}^{2}]$
and $\beta_{a}^{*}=\text{argmin}_{\beta\in\mathbb{R}^{p}}E[\{Y-\mu_{a}(X^{T}\beta_{a})\}^{2}]$,
$a=0,1$. As is common in the empirical literature, we assume  a generalized linear model for the OM model in
Assumption \ref{outcome} and a logistic
regression model for the PS model in Assumption
\ref{PS}. If working models $e(X^{T}\alpha)$ and
$\mu_{a}(X^{T}\beta_{a})$ are correctly specified, we have $e(X)=e(X^{T}\alpha^{*})$
and $\mu_{a}(X)=\mu_{a}(X^{T}\beta_{a}^{*})$, respectively. However, the working
models may be misspecified. $\mathcal{M}_{\alpha}$ denotes the set
of true important variables in the PS model, and $\mathcal{M}_{\beta}$
denotes the set of true important variables in the OM model.
Define the union set of true important variables in the PS and the
OM model as $\mathcal{U}$, i.e., $\mathcal{U}=\mathcal{M}_{\alpha}\cup\mathcal{M}_{\beta}$,
and the intersection set of true important variables in the PS and
the OM model as $\mathcal{I}$, i.e., $\mathcal{I}=\mathcal{M}_{\alpha}\cap\mathcal{M}_{\beta}$.
We refer to true important sets as oracle sets. We use the hat for
the set consisting of the selected variables by variable selection procedures.
The set of variables selected from the PS model is denoted by $\widehat{\mathcal{M}}_{\alpha}$,
and the set of variables selected from the OM model is denoted by
$\widehat{\mathcal{M}}_{\beta}$. Likewise, we define $\widehat{\mathcal{U}}=\widehat{\mathcal{M}}_{\alpha}\cup\widehat{\mathcal{M}}_{\beta}$
and $\widehat{\mathcal{I}}=\widehat{\mathcal{M}}_{\alpha}\cap\widehat{\mathcal{M}}_{\beta}$.

\section{Variable Selection Criteria } \label{3 method}

\subsection{Classification of pretreatment variables }

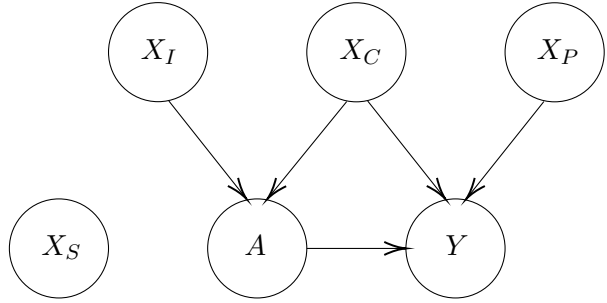
\begin{figure}
\centering 

\begin{tikzpicture}[x=0.75pt,y=0.75pt,yscale=-1,xscale=1]

\draw   (190,106) .. controls (189.96,92.19) and (201.13,80.97) .. (214.93,80.93) .. controls (228.74,80.9) and (239.96,92.06) .. (240,105.87) .. controls (240.04,119.68) and (228.87,130.9) .. (215.07,130.93) .. controls (201.26,130.97) and (190.04,119.81) .. (190,106) -- cycle ;
\draw   (290,106) .. controls (290,92.19) and (301.19,81) .. (315,81) .. controls (328.81,81) and (340,92.19) .. (340,106) .. controls (340,119.81) and (328.81,131) .. (315,131) .. controls (301.19,131) and (290,119.81) .. (290,106) -- cycle ;
\draw   (390,106) .. controls (390,92.19) and (401.19,81) .. (415,81) .. controls (428.81,81) and (440,92.19) .. (440,106) .. controls (440,119.81) and (428.81,131) .. (415,131) .. controls (401.19,131) and (390,119.81) .. (390,106) -- cycle ;
\draw   (240,205) .. controls (240,191.19) and (251.19,180) .. (265,180) .. controls (278.81,180) and (290,191.19) .. (290,205) .. controls (290,218.81) and (278.81,230) .. (265,230) .. controls (251.19,230) and (240,218.81) .. (240,205) -- cycle ;
\draw   (340,205) .. controls (340,191.19) and (351.19,180) .. (365,180) .. controls (378.81,180) and (390,191.19) .. (390,205) .. controls (390,218.81) and (378.81,230) .. (365,230) .. controls (351.19,230) and (340,218.81) .. (340,205) -- cycle ;
\draw    (310.14,130.86) -- (271.41,178.45) ;
\draw [shift={(270.14,180)}, rotate = 309.14] [color={rgb, 255:red, 0; green, 0; blue, 0 }  ][line width=0.75]    (10.93,-3.29) .. controls (6.95,-1.4) and (3.31,-0.3) .. (0,0) .. controls (3.31,0.3) and (6.95,1.4) .. (10.93,3.29)   ;
\draw    (320.71,130.57) -- (358.62,178.43) ;
\draw [shift={(359.86,180)}, rotate = 231.62] [color={rgb, 255:red, 0; green, 0; blue, 0 }  ][line width=0.75]    (10.93,-3.29) .. controls (6.95,-1.4) and (3.31,-0.3) .. (0,0) .. controls (3.31,0.3) and (6.95,1.4) .. (10.93,3.29)   ;
\draw    (409.57,130.86) -- (371.67,178.44) ;
\draw [shift={(370.43,180)}, rotate = 308.53999999999996] [color={rgb, 255:red, 0; green, 0; blue, 0 }  ][line width=0.75]    (10.93,-3.29) .. controls (6.95,-1.4) and (3.31,-0.3) .. (0,0) .. controls (3.31,0.3) and (6.95,1.4) .. (10.93,3.29)   ;
\draw    (220.43,130.57) -- (258.33,178.43) ;
\draw [shift={(259.57,180)}, rotate = 231.62] [color={rgb, 255:red, 0; green, 0; blue, 0 }  ][line width=0.75]    (10.93,-3.29) .. controls (6.95,-1.4) and (3.31,-0.3) .. (0,0) .. controls (3.31,0.3) and (6.95,1.4) .. (10.93,3.29)   ;
\draw    (290,205) -- (338,205) ;
\draw [shift={(340,205)}, rotate = 180] [color={rgb, 255:red, 0; green, 0; blue, 0 }  ][line width=0.75]    (10.93,-3.29) .. controls (6.95,-1.4) and (3.31,-0.3) .. (0,0) .. controls (3.31,0.3) and (6.95,1.4) .. (10.93,3.29)   ;
\draw   (140,205) .. controls (140,191.19) and (151.19,180) .. (165,180) .. controls (178.81,180) and (190,191.19) .. (190,205) .. controls (190,218.81) and (178.81,230) .. (165,230) .. controls (151.19,230) and (140,218.81) .. (140,205) -- cycle ;

\draw (155,196.83) node [anchor=north west][inner sep=0.75pt]    {$X_{S}$};
\draw (205,98.26) node [anchor=north west][inner sep=0.75pt]    {$X_{I}$};
\draw (305.29,98.26) node [anchor=north west][inner sep=0.75pt]    {$X_{C}$};
\draw (404.71,98.26) node [anchor=north west][inner sep=0.75pt]    {$X_{P}$};
\draw (257.29,196.83) node [anchor=north west][inner sep=0.75pt]    {$A$};
\draw (358.71,196.83) node [anchor=north west][inner sep=0.75pt]    {$Y$};

\end{tikzpicture} 
\caption{A diagram of pretreatment covariate structure. Abbreviations: $X_{I}$,
instrumental variables; $X_{C}$, confounding variables; $X_{P}$,
precision variables; $X_{S}$, spurious variables}
\label{fig:diagram}
\end{figure}

In the presence of a considerable number of spurious covariates, including
unnecessary covariates in the model can lead to statistical inefficiency of the estimation or sometimes be computationally infeasible. For this reason, variable selection is essential to exclude unnecessary covariates. We investigate the variable selection approaches for the AIPW estimator of the ACE, given its desirable double robustness property. As mentioned in the
introduction, typically, there are four main types of pretreatment variables: instrumental variables ($X_{I}$), confounder variables ($X_{C}$), precision variables ($X_{P}$), and spurious variables ($X_{S}$). Figure \ref{fig:diagram} displays relationships of pretreatment variables.

Since the pretreatment variables play different roles in the estimation, which variables to be selected depends on the goal for variable selection. We categorize four goals for variable selection: (1) variable selection for prediction modeling; (2) variable selection for confounding control; (3) variable selection for efficient estimation; and (4) variable selection for double robustness. We discuss them in the following subsections.
\subsection{Variable selection for prediction modeling}
Traditionally, variable selection aims to gain prediction accuracy. It is crucial to decide which variables to include in models because the choice of variables strongly influences the model's performance. The exclusion of essential variables from the models fails to identify a genuine relationship between outcomes, treatment, and covariates.  

Backward elimination, forward selection, stepwise selection, $p$ values, Akaike information criterion, Bayesian information criterion, and Mallows' $C_p$ statistic are traditional variable selection procedures \citep{chowdhury2020variable}. These methods can be computationally expensive and do not consider stochastic errors caused in the stages of variable selections \citep{fan2001variable}. \citet{tibshirani1996regression} proposes the least absolute shrinkage and selection operator (LASSO), which is the penalized least squares estimate with the $L_1$ penalty in the squares and likelihood settings. \citet{zou2006adaptive} demonstrate situations where the LASSO selection is inconsistent and present an alternative method, the Adaptive LASSO, where adaptive weights are employed to penalize different coefficients in the $l_1$ penalty. \citet{fan2001variable} show that the LASSO shrinkage gives a biased estimator for the large coefficients and propose the SCAD that takes advantage of the oracle properties. The procedure has the oracle properties if it identifies the right subset model and has the optimal estimation rate. 
The change in estimate criterion method attempts to obtain a low-bias estimator with a minimal covariate set. The method removes a covariate one at a time from a covariate set and then compares the estimates between the reduced set and the original set. If the change in estimate is significant enough, the covariate is removed from the adjustment set. 

However, only focusing on the high prediction accuracy of the PS model or the OM model is not practically helpful for obtaining reasonable ACE estimates. If one cares about only the PS predictor, the AIPW estimator may lose efficiency, excluding $X_P$. We will discuss this in Subsection 3.4. If one cares about only the outcome predictor, the AIPW estimator may no longer retain a double robustness property. We will discuss this in Subsection 3.5.

\subsection{Variable selection for confounding control}
The omission of $X_C$ that appears in both OM and PS models makes the ACE inconsistent. Thus, all $X_C$ should be measured and included for the unbiased ACE. However, identifying all the confounders is not easy in practice. In the real world, the relationship between covariates and the treatment, as well as the relationship between covariates and the outcome, are not fully known. For this reason, a richly parameterized PS model is preferred to ensure the inclusion of $X_C$. In this process, too many covariates are involved in the model, which leads to a complicated model to estimate. An alternative approach is to adjust for covariates that are common causes of exposure and outcome. However, if the information on the common cause of treatment and outcome is not clear, there might be a missed set of covariates which is required to adjust for confounding \citep{vanderweele2019principles}. Besides, this approach does not take into account $X_P$ helpful for efficiency. Thus, this variable selection approach is not suitable for efficient and accurate estimation. 

\subsection{Variable selection for efficient estimation }
One of the primary motivations for variable selection is to gain efficiency in estimating the ACE.
\citet{lunceford2004stratification} show that when including $X_{P}$
as well as $X_{C}$ into the PS model, all weighted estimators for
the ACE are consistent, and the variance of the AIPW estimator based on $X_{P}$
and $X_{C}$ has a smaller variance relative to the one based only
on $X_{C}$. \citet{brookhart2006variable} suggest using
only outcome predictors for the PS model since $X_{P}$ reduces
the variance while $X_{I}$ and $X_{S}$ inflate the variance without
changing bias. \citet{henckel2019graphical} introduce a new graphical criterion. Using their Theorem 3.1, it is shown that the outcome predictor-based asymptotic variance is smaller than the treatment predictor-based asymptotic variance.  \citet{rotnitzky2019efficient} extend the results to non-parametric causal graphical models. In their settings, $X_C$ affects outcomes through $X_P$. They point out that the stronger association between $X_C$ and the PS model and the weaker association between $X_C$ and $X_P$ are, the less efficient it is to include $X_C$. \citet{tang2020outcome} also show that adjusting for outcome predictors improves efficiency, and adjusting for $X_I$ increases the variance. To show these results, they quantify the difference in the asymptotic variance of the estimators based on the $X_P$ and $X_I$. 

Many methods have been developed to obtain efficiency, including outcome predictors and excluding $X_I$. 
\citet{shortreed2017outcome} derive the outcome-adaptive lasso.
\citet{ertefaie2018variable} propose a penalized objective function, which selects outcome predictors while excluding $X_I$ and $X_S$.
\citet{henckel2019graphical} establish a procedure to prune a valid adjustment to get a valid subset with a smaller asymptotic variance. \citet{tang2020outcome} propose the causal ball screening for selecting all outcome predictors from modern ultra-high dimensional data sets and excluding $X_I$ and $X_S$.

\subsection{Variable selection for double robustness }

As mentioned in the previous subsection, outcome predictor approaches make the ACE estimator more efficient under the correct models. However, there is a lack of information on the relationship between covariates or on both model specifications in the real world. The insufficient information makes it difficult to obtain the correct models and to identify relevant covariates that increase the efficiency of the ACE estimator.
In this case, a robust estimation can be more reliable, although a bit of efficiency is sacrificed. In particular, this is obvious when one does not have complete knowledge of the OM model and uses the AIPW estimator for the ACE estimation.
Suppose linear models are
assumed for the non-linear OM models. Then, no variables may
be selected, or some variables that are not associated with the true outcomes
may be chosen for the OM model. In this case, including $X_{I}$
can be helpful in terms of robustness. To illustrate, consider the
following example. 

  \begin{example}\label{example 1}
Let $X=(X_{1},X_{2},X_{3})$ be
the vector of pretreatment covariates. Assume the true OM model is non-linear, e.g., $\mu_{a}(X)=0.1X_{1}^{2}+X_{2}^{2}+2X_{2}$; the true PS model follows a logistic model, e.g.,
$\text{logit}\{e(X)\}=1+X_{1}+X_{3}$. Then, $\mathcal{M}_{\beta}=\{X_{1},X_{2}\}$
and $\mathcal{M}_{\alpha}=\{X_{1},X_{3}\}$. Thus, $X_{1} \in X_C$,
$X_{2} \in X_P$, and $X_{3} \in X_I$. Assume treatment assignment
follows a correct logistic regression model, and the OM model may
be misspecified as linear. In this case, since $X_{1}$ is non-linearly related to the OM models, $X_{1}$ is unlikely to be
selected for the OM model in most cases. Consequently, we have
$\widehat{\mathcal{M}}_{\beta}=\{X_{2}\}$ and $\widehat{\mathcal{M}}_{\alpha}=\{X_{1},X_{3}\}$.
  \end{example}

In Example \ref{example 1}, a linear model is assumed for the non-linear
OM model, whereas the PS model is correctly specified. That
is, Assumption \ref{outcome} is not satisfied and $\mu(X,\beta_{a}^{*})\ne\mu(X)=E(Y|X)$.
On the other hand, Assumption \ref{PS} is satisfied and $e(X,\alpha^{*})=e(X)=E(A|X)$.
Generally, the AIPW estimator should be consistent with the ACE by its doubly
robust property because Assumption \ref{PS} is satisfied. However, in Example \ref{example 1}, we cannot obtain the consistent PS model with the outcome predictor
approach because $X_C=\{X_1\}$ is likely to be dropped due to its weak linear relationship and $\widehat{\mathcal{M}}_{\beta}=\{X_{2}\}$ is not involved with the PS model. 
Thus, both the PS and OM models using a set based on the outcome predictors are not consistent. As a result, the AIPW estimator loses its doubly robust property, and it is no longer consistent for ACE. However, if $\widehat{\mathcal{U}}=\widehat{\mathcal{M}}_\alpha \cup \widehat{\mathcal{M}}_\beta=\{X_1, X_2, X_3\}$ is used to estimate the PS model, we can obtain the consistent PS model, and the AIPW estimator is consistent with ACE. Although $X_I=X_1$ may reduce efficiency, it is necessary to get a consistent AIPW estimator.
 Therefore, we suggest using $\widehat{\mathcal{U}}$
to retain the double robustness of the AIPW estimator. In the simulation
study, we will show that the AIPW estimator based on $\widehat{\mathcal{I}}$
or $\widehat{\mathcal{M}}_{\beta}$ is not doubly robust when the PS
model is correctly specified, but the OM model is misspecified. 

\subsection{Proposed procedure for variable selection and estimation }

The proposed method has three steps: we separately select important
variables for the OM model in Step 1 and the PS
model in Step 2. In Step 3, we estimate the ACE using a doubly robust
estimator based on the union set of selected covariates in Step 1
and Step 2. In Step 1 and Step 2, we employ a penalized estimating
equation for determining important covariates. The SCAD is used here. We specify $q_{\lambda}(x)$
to be a folded concave SCAD
penalty function \citep{fan2011nonconcave, yang2020doubly}. The SCAD penalty is defined
by 
\begin{equation}\label{eq:q}
q_{\lambda}(|\theta|)=\lambda\bigg\{ I(|\theta|<\lambda)+\frac{(a\lambda-|\theta|)_{+}}{(a-1)\lambda}I(|\theta|\ge\lambda)\bigg\}
\end{equation}
for $a>0$, where $(\cdot)_{+}$ is the truncated linear function;
i.e., if $x\ge0$, $(x)_{+}=x$, and if $x<0$, $(x)_{+}=0$. We use
$a=3.7$ following the suggestion of \cite{fan2001variable}.

In Step 1, we run the SCAD for each of the OM models
separately. For $\beta_{1}$, we conduct the SCAD only using the observations
$(Y_{i},X_{i})$ with $A_{i}=1$. Likewise, the observations $(Y_{i},X_{i})$
with $A_{i}=0$ are used for $\beta_{0}$. The penalized estimating
functions for $\beta_{a}$, $a=0,1$, are defined as 
\begin{equation}
U_{1}(\beta_{1})=\frac{1}{n_{1}}\sum_{i=1}^{n}A_{i}\{Y_{i}-\mu_{1}(X_{i}^{T}\beta_1)\}X_{i}-q_{\lambda_{\beta_1}}(|\beta_1|)\cdot\text{sign}(\beta_{1}),\label{eq: beta1}
\end{equation}
\begin{equation}\label{eq: beta0}
U_{2}(\beta_{0})=\frac{1}{n_{0}}\sum_{i=1}^{n}(1-A_{i})\{Y_{i}-\mu_{0}(X_{i}^{T}\beta_0)\}X_{i}-q_{\lambda_{\beta_0}}(|\beta_0|)\cdot\text{sign}(\beta_{0}),
\end{equation}
where $n_{1}=\sum_{i=1}^{n}I(A_{i}=1)$ and $n_{0}=\sum_{i=1}^{n}I(A_{i}=0)$. $q_{\lambda_{\beta_a}}(|\beta_a|)$ is defined in Eq. \eqref{eq:q}.

In Step 2, we implement the SCAD for the PS model. The corresponding
penalized estimating function for $\alpha$ is 
\begin{equation}
U_{3}(\alpha)=\frac{1}{n}\sum_{i=1}^{n}\{A_{i}-e(X_{i};\alpha)\}X_{i}-q_{\lambda_{\alpha}}(|\alpha|)\cdot\text{sign}(\alpha).\label{eq: alpha}
\end{equation}
Let $(\tilde{\alpha},\tilde{\beta}_{0},\tilde{\beta}_{1})$ denote
the solution for the penalized joint estimating equation $U=0$. In
this procedure, $\widehat{\mathcal{M}}_{\alpha}$ is the set of variables
that correspond to the nonzero coefficients in the PS model, and $\widehat{\mathcal{M}}_{\beta}$
is the set of variables that correspond to the nonzero coefficients
in the OM model. Likewise, $\widehat{\mathcal{U}}=\widehat{\mathcal{M}}_{\alpha}\cup\widehat{\mathcal{M}}_{\beta}$
and $\widehat{\mathcal{I}}=\widehat{\mathcal{M}}_{\alpha}\cap\widehat{\mathcal{M}}_{\beta}$.
\citet{fan2001variable} show that the SCAD estimators perform the
oracle procedure in variable selection, which means they behave as
if the correct submodels were known. Thus, the set $\widehat{\mathcal{U}}$
includes the true important variables in either the PS model or the
OM model with probability approaching one.

In Step 3, we re-estimated the coefficients using the variables in
the set $\widehat{\mathcal{U}}$ and then derive the AIPW estimator of
the ACE. At this time, we use estimating functions without  penalty. Under Assumption \ref{outcome}, for the OM model,
let 
\begin{equation} \label{eq: est_beta}
S_{a}(X,Y;\beta_{a})=\frac{\partial\mu_{a}(X;\beta_{a})}{\partial\beta_{a}}\{Y-\mu_{a}(X;\beta_{a})\}
\end{equation}
be the estimating function for $\beta_{a}^{*}$ for $a=0,1$. Under
Assumption \ref{PS}, for the PS model, let 
\begin{equation} \label{eq: est_alpha}
S(A,X;\alpha)=\frac{A-e(X;\alpha)}{e(X;\alpha)\{1-e(X;\alpha)\}}\frac{\partial e(X;\alpha)}{\partial\alpha}
\end{equation}
be the estimating function for $\alpha^{*}$. $\widehat{U}^{C}$ is defined as the complement of $\widehat{U}$. We use the estimating equations (Eq. \ref{eq: est_beta} and \ref{eq: est_alpha}) restricted to the parameter space $\{\beta_a: \beta_{a,\hat{\mathcal{U}}^C}=0 \}$ and $\{\alpha: \alpha_{\hat{\mathcal{U}}^C}=0 \}$, respectively.
Let $(\hat{\alpha}, \hat{\beta}_0, \hat{\beta}_1)$ denote the solution for Eq. \eqref{eq: est_beta} and \eqref{eq: est_alpha}.
To summarize, our two-stage procedure for variable selection and estimation
is as follows. 
\begin{itemize}
\item Step 1:  Under Assumption \ref{outcome}, use the penalization method to
select important variables in the OM model using the SCAD, i.e., $\widehat{\mathcal{M}}_{\beta}=\{j:\Tilde{\beta}_{a,j}\ne0\}$,
where $\tilde{\beta}_{a,j}$ is the solution for Eq. \eqref{eq: beta1}
and \eqref{eq: beta0}, for $a=0,1$. 
\item Step 2:  Under Assumption \ref{PS}, use the penalization method to select
important variables in the PS model using the SCAD, i.e., $\widehat{\mathcal{M}}_{\alpha}=\{j:\Tilde{\alpha}_{j}\ne0\}$,
where $\Tilde{\alpha}_{j}$ is the solution for Eq. \eqref{eq: alpha}. 
\item Step 3:  Let the set of variables for estimation be $\widehat{\mathcal{U}}=\widehat{\mathcal{M}}_{\alpha}\cup\widehat{\mathcal{M}}_{\beta}$.
The proposed estimator is 
\begin{align}
&\aipwhat\label{eq:AIPW}\\
=&\frac{1}{n}\sum_{i=1}^{n}\Bigg[\frac{A_{i}Y_{i}}{e(X_{i};\hat{\alpha})}+\bigg\{1-\frac{A_{i}}{e(X_{i};\hat{\alpha})}\bigg\}\mu_{1}(X_{i};\hat{\beta}_{1})\nonumber\\ 
-&\frac{(1-A_{i})Y_{i}}{1-e(X_{i};\hat{\alpha})}-\bigg\{1-\frac{(1-A_{i})Y_{i}}{1-e(X_{i};\hat{\alpha})}\bigg\}\mu_{0}(X_{i};\hat{\beta}_{0})\Bigg], \nonumber
\end{align}
where $\hat{\alpha}$ and $\hat{\beta}_a$ are obtained by fitting the OM and PS models for $\alpha$ and $\beta$ with $X_i \in \widehat{\mathcal{U}}, i=1,...,n$.
\end{itemize}
In Steps 1 and 2, the choice of the
regularization parameter $\lambda$ is crucial because it controls
the model's sparsity level. In many research, $\lambda$ is chosen
by cross-validation. However, according to \cite{meinshausen2010stability},
$\lambda$ chosen from cross-validation selects too many noise variables
in a high-dimensional setting. We modify the R function \texttt{cv.ncvreg} in the \texttt{ncvreg} package so that cross-validation selects the regularization parameter
($\lambda_{a}$, $\lambda_{b}$) from a pre-range of $\lambda$. In
this way, we can prevent  over-selecting. \texttt{cv.ncvreg}
solves the estimating function using a coordinate descent algorithm.
The coordinate descent algorithms minimize the target function with
respect to a single parameter at a time, with other components of
the variable vector $X$ being fixed at their current values. If specifying ($\lambda_a$, $\lambda_b$) is difficult, another approach for resolving the overselecting problem is to utilize other variable selection methods, such as the Adaptive LASSO \citep{zou2006adaptive}. See Section S3 in the supplementary material.



\section{Asymptotic Results for Variable Selection and Estimation } \label{4 asymptotic}

Under certain regularity conditions given in \cite{fan2001variable},
$\tilde{\alpha}$ and $\tilde{\beta}_{a}$ satisfy the
selection consistency and the oracle properties under penalized likelihood
for both linear regression and logistic regression. Hence, we can obtain
$\Vert\tilde{\alpha}-\alpha^*\Vert_{2}=O_{p}\{(p/n)^{1/2}\}$
and $\Vert\tilde{\beta}_{a}-\beta_{a}^*\Vert_{2}=O_{p}\{(p/n)^{1/2}\}$,
for $a=1,2$. Now we focus on the asymptotic behavior of the AIPW
estimator based on $(\hat{\alpha},\hat{\beta}_{0},\hat{\beta}_{1})$, which are obtained by fitting the OM and PS models.
We consider an influence function to study the asymptotic properties
of the proposed estimator. Under mild regularity conditions 
\citep[e.g.,][]{robins1994estimation}, 
\begin{equation*}
\aipwst-\tau=\frac{1}{n}\sum_{i=1}^{n}\psi(A_{i},X_{i},Y_{i})+o_{p}(1),
\end{equation*}
where $\psi(A,X,Y)$ is the influence function of $\hat{\tau}$ with
$E(\psi)=0$ and $E(\psi^{2})<\infty$ \citep {bickel1993efficient}.

Let 
\begin{align*}
\Sigma_{\alpha}&=E\left\{ S^{\otimes2}(A,X;\alpha)\right\} \\&=E\left[\frac{1}{e(X;\alpha^{*})\{1-e(X;\alpha^{*})\}}\left\{ \frac{\partial e(X;\alpha^{*})}{\partial\alpha}\right\} ^{\otimes2}\right]
\end{align*}
be the Fisher information matrix for $\alpha$ in the PS model. For
simplicity, denote 
\begin{align*}
&e_{i}^{*}=e(X_{i};\alpha^{*}),\\ 
&\dot{e}_{i}^{*}=\partial e(X_{i};\alpha^{*})/\partial\alpha^{T},\\ 
&S_{i}^{*}=S(A_{i},X_{i};\alpha^{*}),
\\&\mu_{ai}^{*}=\mu_{a}(X_{i};\beta_{a}^{*}),\\
&\dot{\mu}_{ai}^{*}=\partial\mu_{a}(X_{i};\beta_{a}^{*})/\partial\beta_{a}^{T},\\
&S_{ai}^{*}=S_{a}(A_{i},X_{i},Y_{i};\beta_{a}^{*}),\\
&\dot{S}_{ai}^{*}=\partial S_{a}(A_{i},X_{i},Y_{i};\beta_{a}^{*})/\partial\beta_{a}^{T},
\end{align*}
for $a=0,1$. Under Assumption \ref{outcome} or Assumption \ref{PS},
the influence function for $\aipwhat$ can be written as following:
\begin{align}
\begin{split}&\psi(A_{i},X_{i},Y_{i})\\ & =\frac{A_{i}Y_{i}}{e_{i}^{*}}+\left(1-\frac{A_{i}}{e_{i}^{*}}\right)\mu_{1i}^{*}-\frac{(1-A_{i})Y_{i}}{1-e_{i}^{*}}-\left(1-\frac{1-A_{i}}{1-e_{i}^{*}}\right)\mu_{0i}^{*}\\
 & +E\bigg[\bigg\{\frac{A(Y-\mu_{1}^{*})}{(e^{*})^{2}}+\frac{(1-A)(Y-\mu_{0}^{*})}{(1-e^{*})^{2}}\bigg\}\dot{e}^{*}\bigg]\Sigma_{\alpha}^{-1}S_{i}^{*}\\
 & -E\bigg\{\bigg(1-\frac{A}{e^{*}}\bigg)\dot{\mu}_{1}^{*}\bigg\}\Big\{ E(\dot{S}_{1}^{*})\Big\}^{-1}S_{1i}^{*}\\
 & +E\bigg\{\bigg(1-\frac{1-A}{1-e^{*}}\bigg)\dot{\mu}_{0}^{*}\bigg\}\Big\{ E(\dot{S}_{0}^{*})\Big\}^{-1}S_{0i}^{*}-\tau.
\end{split}
\label{eq: TE of IF}
\end{align}
The regularity condition for the SCAD and the details of Eq. \eqref{eq: TE of IF}
are presented in the supplementary material.  Note that if $e^{*}$ and
$\mu_{a}^{*}$ is correctly specified, $e^{*}=e(X;\alpha^{*})=e(X)$
and $\mu_{a}^{*}=\mu_{a}(X;\beta_{a}^{*})=\mu_{a}(X)$, for $a=1,2$.
Hence, if $e^{*}$ or $\mu_{a}^{*}$ is the correct model, the right-hand
side of Eq. \eqref{eq: TE of IF} is zero, as shown in the supplementary material. It follows that $E(\psi)=0$ and $E(\psi\psi^{T})<\infty$.
Thus, we have 
\begin{equation*}
\aipwhat-\tau=\frac{1}{n}\sum_{i=1}^{n}\psi(A_{i},X_{i},Y_{i})+o_{p}(1),
\end{equation*}
As a result, $\sqrt{n}\{\aipwhat-\tau\}$ is also asymptotically normal.

\begin{theorem} {}
Under Assumptions 1-4 and the regularity conditions specified in the supplementary material, if either $e(X^T \alpha)$ or $m(X^T \beta)$ is correctly specified,
\begin{align}\label{eq: asym_normal}
    \sqrt{n}\{\aipwhat-\tau\} \rightarrow \mathcal{N}\{0, E(\psi \psi^T)\},
\end{align}
in distribution, as $n \to \infty$, where $\psi$ is defined in Eq. (\ref{eq: TE of IF}).
\end{theorem}
 Additionally, when the propensity score and the regression function are correctly specified,  ${\hat{\tau}}$ achieves the semiparametric efficiency bound \citep{glynn2010introduction}.

\begin{theorem} [Double robustness of the proposed estimator] Under Assumption 1-2, if either Assumption 3 or Assumption 4 holds, not necessarily both,  $\aipwhat$ in Equation \eqref{eq:AIPW} is consistent with $\tau$.
\end{theorem}
The proof is given in the supplementary material.

There are two approaches to estimating the asymptotic variance: (a)
conducting the bootstrap and (b) using the point estimation for the
asymptotic variance, replacing $\psi$ with $\widehat{\psi}$ in (\ref{eq: asym_normal}).
For the former, since $\hat{\tau}$ is asymptotically linear and normal,
we can estimate the valid variance estimator of $\hat{\tau}$ by bootstrapping
original observations. For the latter, we use $\widehat{\psi}$ as an
estimator for $\psi$ in Eq. \eqref{eq: TE of IF}, substituting $(\hat{\alpha},\hat{\beta}_{0},\hat{\beta}_{1})$
for $(\alpha^{*},\beta_{0}^{*},\beta_{1}^{*})$. Then, we can estimate
the asymptotic variance of $\widehat{\tau}_{n}(\hat{\alpha},\hat{\beta}_{0},\hat{\beta}_{1})$,
$V\{\hat{\tau}_{n}(\hat{\alpha},\hat{\beta}_{0},\hat{\beta}_{1})\}$,
by 
\begin{align*}
\widehat{V}\{\hat{\tau}_{n}(\hat{\alpha},\hat{\beta}_{0},\hat{\beta}_{1})\}=\frac{1}{n}\widehat E(\widehat{\psi}\widehat{\psi}^{T})=\frac{1}{n^{2}}\sum_{i=1}^{n}\widehat{\psi}\widehat{\psi}^{T}.
\end{align*}
Considering the long time it takes to conduct bootstrap, we use the latter in
Sections \ref{5 simulation} and \ref{6 application}.

\section{Simulation Study}\label{5 simulation}

In this section, we conduct a simulation study to evaluate the finite
sample performances of the doubly robust ACE estimator with different
variable selection strategies. Additionally, the simulation study is done under model misspecification to manifest the robustness of the proposed estimator.

\subsection{Simulation setup}

We generate the dataset with size $n=5000$. The covariate $X_{i}=(1,X_{1,i},...,X_{p-1,i})^{T}$
is $p$-dimensional, where $p$ is set to be $50$. The first component
is one, and the others are independently generated from the standard
normal with mean 0 and variance 1. Table \ref{tab:scenario} summarizes
the structure of pretreatment variables for four scenarios. In all scenarios, the
last $p-6$ coefficients were set to 0 in the PS and OM models,
representing $p-6$ spurious covariates, and $X_C=\{X_{3,i}, X_{4,i}\}$. In Scenarios 1 and 4, $X_I=\{X_{1,i}, X_{2,i}\}$. In Scenarios 1 and 3, $X_P=\{X_{5,i}, X_{6,i}
\}$.

\begin{table}
\centering
\caption{\label{tab:scenario}The pretreatment variables in four scenarios}
\begin{tabular}{cccc}
\toprule
 & $X_{I}$ & $X_{C}$ & $X_{P}$\tabularnewline
\midrule
Scenario 1 & $X_{1,i}$, $X_{2,i}$ & $X_{3,i}$, $X_{4,i}$ & $X_{5,i}$, $X_{6,i}$\tabularnewline
Scenario 2 &  & $X_{3,i}$, $X_{4,i}$ & \tabularnewline
Scenario 3 &  & $X_{3,i}$, $X_{4,i}$ & $X_{5,i}$, $X_{6,i}$\tabularnewline
Scenario 4 & $X_{1,i}$, $X_{2,i}$ & $X_{3,i}$, $X_{4,i}$ & \tabularnewline
\bottomrule
\end{tabular}
\end{table}

We generate a binary treatment, $A_{i}$, from a Bernoulli distribution
with the PS. For the PS model, we consider both a linear
model (PSM I) and a non-linear model (PSM II):
\begin{itemize} 
\item PSM I:  $\text{logit}(e_{i})=\alpha_{1}^{T}X_{i},$ 
\item PSM II: $\text{logit}(e_{i})=3.5+\alpha_{2}^{T}\log(X_{i}^{2})-\cos(X_{3,i}+X_{4,i})$, 
\end{itemize}
where $\alpha_{1}$ and $\alpha_{2}$ are $(p-1)$-dimensional vectors
of coefficients in the PS model. The true values of $\alpha_{1}$
and $\alpha_{2}$ are defined differently depending on the scenario
structure: 
\begin{itemize} 
\item Scenario 1:  $\alpha_{1}=(0,1,1,1,1,0,...,0)^{T}$ and $\alpha_{2}=(0,3,3,3,3,0,...,0)^{T}$, 
\item Scenario 2:  $\alpha_{1}=(0,0,0,1,1,0,...,0)^{T}$ and $\alpha_{2}=(0,0,0,3,3,0,...,0)^{T}$, 
\item Scenario 3:  $\alpha_{1}=(0,0,0,1,1,0,...,0)^{T}$ and $\alpha_{2}=(0,0,0,3,3,0,...,0)^{T}$, 
\item Scenario 4:  $\alpha_{1}=(0,1,1,1,1,0,...,0)^{T}$ and $\alpha_{2}=(0,3,3,3,3,0,...,0)^{T}$. 
\end{itemize}
For generating continuous outcome variable, $Y_{i}$, we consider
both linear (OM I) and non-linear OM models (OM II): 

\begin{itemize} 
\item OM I:    $Y_{i}=\beta_{a}^{T}X_{i}+\epsilon_{i}$, $\epsilon_{i}\sim\mathcal{N}(0,1)$,
where $a=0,1$, 
\item OM II: \\  $Y_{0}=1+\exp(\sin(\beta_{0}^{T}X_{i}))-2\cos(\beta_{0,4}X_{3,i}+\beta_{0,5}X_{4,i})+\beta_{0,6}X_{5,i}-\beta_{0,7}X_{6,i}+\epsilon_{0,i}$,
$\epsilon_{0,i}\sim\mathcal{N}(0,1)$, \\
 $Y_{1}=1+\exp(2\sin(\beta_{1}^{T}X_{i}))-\cos(\beta_{1,4}X_{3,i}+\beta_{1,5}X_{4,i})+\beta_{1,6}X_{5,i}-\beta_{1,7}X_{6,i}+\epsilon_{1,i}$,
$\epsilon_{1,i}\sim\mathcal{N}(0,1)$, 
\end{itemize}
where $\beta_{0}$ and $\beta_{1}$ are $(p-1)$-dimensional vectors
of coefficients in the OM model, and $\beta_{a,j}$ is the $j$-th
coefficient of $\beta_{a}$ ($a=0,1$). The true values of $\beta_{0}$
and $\beta_{1}$ are defined differently depending on the scenario
structure: 
\begin{itemize} 
\item Scenario 1: 
 $\beta_{0}=(1,0,0,1,1,1,1,...,0)^{T}$ and $\beta_{1}=(1,0,0,2,2,2,2,...,0)^{T}$, 
\item Scenario 2:  $\beta_{0}=(1,0,0,1,1,0,0,...,0)^{T}$ and $\beta_{1}=(1,0,0,2,2,0,0,...,0)^{T}$, 
\item Scenario 3:  $\beta_{0}=(1,0,0,1,1,1,1,...,0)^{T}$ and $\beta_{1}=(1,0,0,2,2,2,2,...,0)^{T}$, 
\item Scenario 4:  $\beta_{0}=(1,0,0,1,1,0,0,...,0)^{T}$ and $\beta_{1}=(1,0,0,2,2,0,0,...,0)^{T}$. 
\end{itemize}
Using the different $\alpha$ and $\beta$ for each Scenario is to make the true important covariates different according to Scenarios. For example, in Scenario 1, $X_{1,i}, X_{2,i}, X_{3,i}$, and  $X_{4,i}$ are important covariates in the PS models and $X_{3,i}, X_{4,i}, X_{5,i}$, and $X_{6,i}$ are in the OM models. In doing so, we can perform the simulations using different combinations of $X_I$, $X_C$, and $X_P$ as in Table \ref{tab:scenario}.
Under the above data-generating mechanisms, the true ACE is zero for
OM I. For OM II, the true ACE is $1.6031$ in Scenarios 1 and 3 and
$1.4280$ in Scenarios 2 and 4.

For each scenario specified in Table \ref{tab:scenario},
we consider four settings: 
\begin{itemize} 
\item Setting (a):  a linear PS model (PSM I) + a linear OM model (OM I), 
\item Setting (b):  a non-linear PS model (PSM II) + a linear OM model (OM I), 
\item Setting (c):  a linear PS model (PSM I) + a non-linear OM model (OM II), 
\item Setting (d):  a non-linear PS model (PSM II) + a non-linear OM model (OM
II). 
\end{itemize}
Although in case the data are generated from the non-linear PS or OM models, we estimate the coefficients based on a linear model. Thus, the PS model in Setting (b), the OM model in Setting (c), and both the PS and OM model in Setting (d) are misspecified. 
In the variable selection steps, the regularization parameter is chosen
by cross-validation. To avoid over-selecting, we select $\lambda$
from $\lambda_{min}$ to $\lambda_{max}$. We set $\lambda_{min}=0.1,0.3,$
and $0.02$ for the linear OM model, the non-linear OM
model, and the PS model, respectively. We set $\lambda_{max}$ to
be the default value provided by the package \texttt{ncvreg}, which is the largest eigenvalue of the design matrix. In this simulation, we use the same set for the PS model and the OM model in the AIPW estimation. Our proposed estimator is the AIPW estimator using the PS and OM models adjusted by $\widehat{\mathcal{U}}$. We compare the performance of our suggested estimator with the AIPW estimator using the PS and OM models adjusted by  $\widehat{\mathcal{I}}$ and $\widehat{\mathcal{M}}_\beta$. Also, our suggested estimator is compared with oracle sets, $\mathcal{U}$, $\mathcal{I}$ and $\mathcal{M}_\beta$, which are the sets consisting of covariates used in the true models.   
We consider the following
estimators: 
\begin{itemize}
\item \texttt{TRUE}: $E(Y_{1}-Y_{0})$, 
\item \texttt{O-UNI}: the AIPW estimator based on $\mathcal{U}$ for comparison
purpose, 
\item \texttt{O-INT}: the AIPW estimator based on $\mathcal{I}$ for comparison
purpose, 
\item \texttt{O-OUT}: the AIPW estimator based on $\mathcal{M}_{\beta}$ for comparison purpose, 
\item \texttt{UNI}: the AIPW estimator based on $\widehat{\mathcal{U}}$ selected
by the SCAD (the proposed approach), 
\item \texttt{INT}: the AIPW estimator based on $\widehat{\mathcal{I}}$ selected
by the SCAD (the confounder-only approach), 
\item \texttt{OUT}: the AIPW estimator based on $\widehat{\mathcal{M}}_{\beta}$
selected by the SCAD (the outcome predictor approach). 
\end{itemize}
Each simulation is based on 2000 Monte Carlo runs. We compute the proportion of over-selecting, under-selecting, average false negatives (the average number of not selected covariates that have true nonzero coefficients), and the average false positives (the average number of selected covariates that have true zero coefficients) for each simulation. We estimate the ACE and obtain the coverage rates of the 95$\%$ confidence interval. 

\subsection{Simulation results}
Table \ref{tab: selection} summarizes the selection performance of the proposed penalization procedure for all scenarios in terms of the proportion of over-selecting (Over), under-selecting (Under), average false negative (FN), and average false positives (FP). The under-selecting proportions for the proposed method are all zeros under the true model specification, implying that most of the true nonzero coefficients are selected by Steps I and II in the proposed procedure.

\renewcommand{\arraystretch}{0.8}
\begin{table*}
\begin{center}
\caption{Simulation results for the selection performance for the proposed penalization procedure}
\label{tab: selection} %
\begin{tabular*}{\textwidth}{@{\extracolsep\fill}lrrrrrrrr@{\extracolsep\fill}}
\toprule
 & \multicolumn{4}{c}{$\beta^{*}$} & \multicolumn{4}{c}{$\alpha^{*}$}\\
 & %
\begin{tabular}{@{}l@{}}
Over\\
$(\times10^{2})$\\
\end{tabular} & %
\begin{tabular}{@{}l@{}}
Under\\
$(\times10^{2})$\\
\end{tabular} & FN & FP & %
\begin{tabular}{@{}l@{}}
Over\\
$(\times10^{2})$\\
\end{tabular} & %
\begin{tabular}{@{}l@{}}
Under\\
$(\times10^{2})$\\
\end{tabular} & FN & FP\\
\hline
 \multicolumn{9}{c}{Scenario 1} \\ \hline
(a) OM I and PSM I & 0.05 & 0 & 0 & 0.0005 & 10.4 & 0 & 0 & 0.1145\\
(b) OM I and PSM II & 4.25 & 0 & 0 & 0.0475 & 12.3 & 100 & 4 & 0.14\\
(c) OM II and PSM I & 0 & 100 & 2 & 0 & 10.4 & 0 & 0 & 0.1145\\
(d) OM II and PSM II & 0.05 & 100 & 2 & 0.0005 & 12.3 & 100 & 4 & 0.14\\
\hline
 \multicolumn{9}{c}{Scenario 2} \\ \hline
(a) OM I and PSM I & 0 & 0 & 0 & 0 & 16.95 & 0 & 0 & 0.2355 \\
(b) OM I and PSM II & 0 & 0 & 0 & 0 & 29.7 & 99.8 & 2 & 0.5075 \\
(c) OM II and PSM I & 0 & 100 & 2 & 0 & 16.95 & 0 & 0 & 0.2355 \\
(d) OM II and PSM II & 0 & 100 & 2 & 0 & 29.7 & 99.8 & 2 & 0.5075 \\
\hline
 \multicolumn{9}{c}{Scenario 3} \\ \midrule
(a) OM I and PSM I & 0 & 0 & 0 & 0 & 16.95 & 0 & 0 & 0.2355 \\
(b) OM I and PSM II & 0 & 0 & 0 & 0 & 29.7 & 99.8 & 2 & 0.5075 \\
(c) OM II and PSM I & 0 & 99.65 & 2 & 0 & 16.95 & 0 & 0 & 0.2355 \\
(d) OM II and PSM II & 0 & 100 & 2 & 0 & 29.7 & 99.8 & 2 & 0.5075 \\
\hline
 \multicolumn{9}{c}{Scenario 4}\\ \midrule
(a) OM I and PSM I & 0.05 & 0 & 0 & 0.0005 & 10.4 & 0 & 0 & 0.1145 \\
(b) OM I and PSM II & 3.75 & 0 & 0 & 0.042 & 12.3 & 100 & 4 & 0.14 \\
(c) OM II and PSM I & 0 & 100 & 2 & 0 & 10.4 & 0 & 0 & 0.1145 \\
(d) OM II and PSM II & 0 & 100 & 2 & 0 & 12.3 & 100 & 4 & 0.14 \\
\hline
\end{tabular*}
\end{center}
\emph{Note}: $X_I$ exists in Scenarios 1 and 4, $X_C$ exists in all Scenarios, and $X_P$ exists in Scenarios 1 and 3. Under OM I (II), the OM model is correctly specified (misspecified), and under PSM I (II), the PS model is correctly specified (misspecified). The results include the proportion of over-selecting, under-selecting, average false negatives, and average false positives for each setting.
\end{table*}

\begin{figure*}
\centering
\includegraphics[width=400pt, height=40pc]{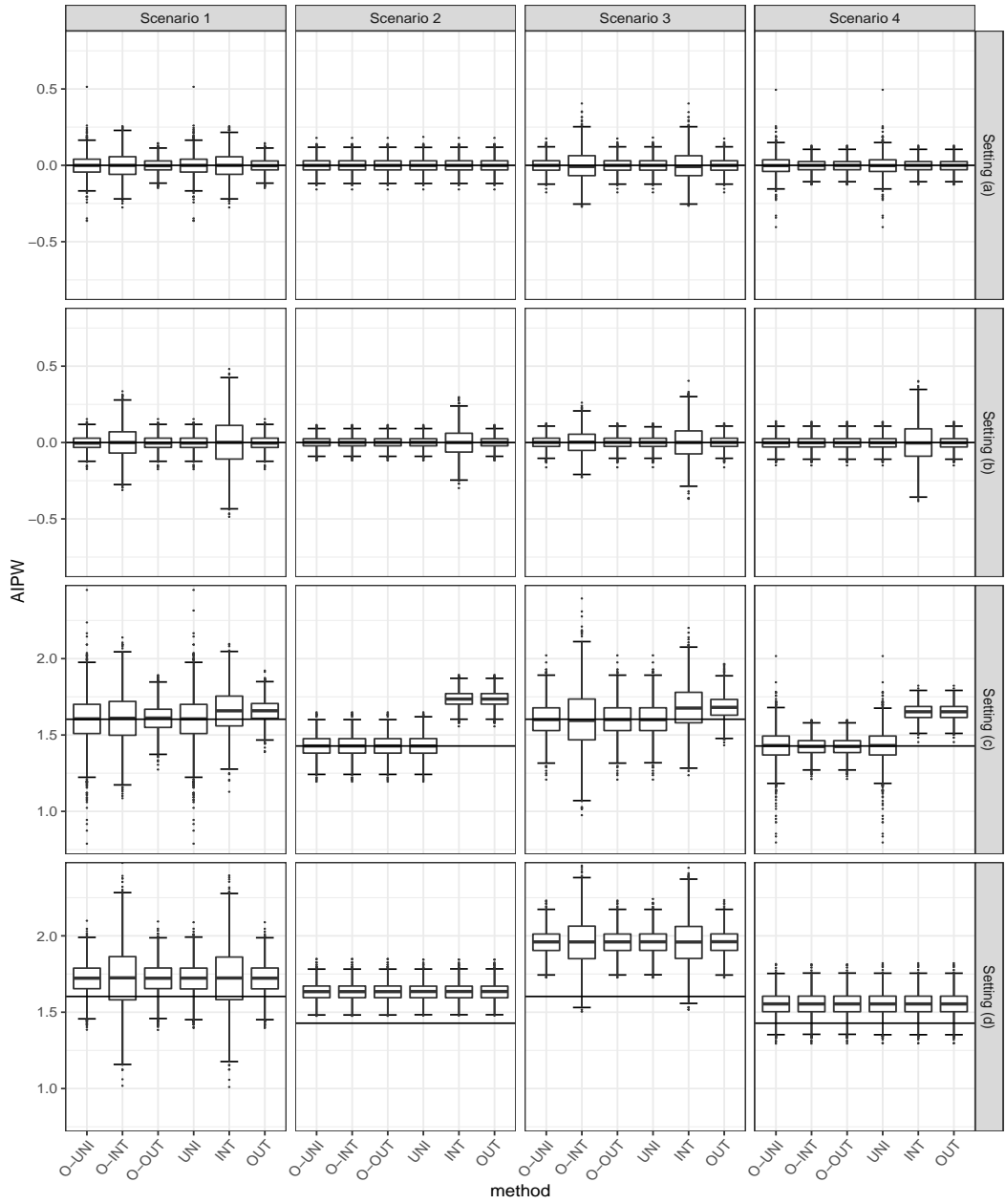}
    \caption{Estimation results under four Scenarios. $X_I$ exists in Scenarios 1 and 4, $X_C$ exists in all Scenarios, and $X_P$ exists in Scenarios 1 and 3. The OM model is correctly specified (misspecified) in Settings (a) and (b) (Settings (c) and (d)), and the PS model is correctly specified (misspecified) in Settings (a) and (c) (Setting (b) and (d)). 
    \label{fig: boxplot}}
\end{figure*}

Figure \ref{fig: boxplot} displays the distribution of the ACE estimates
for all scenarios. $\texttt{\textsc{UNI}}$ and \mtt{OUT} maintain
the efficiency as much as their oracle estimators, while \texttt{\textsc{INT}}
is more variable than its oracle estimator under the misspecification. All methods are more unstable when the OM model is
misspecified than when it is correct. \texttt{\textsc{O-UNI}} and
\mtt{UNI} have more considerable variability than \texttt{\textsc{O-OUT}}
and \mtt{OUT} in Setting (a) and (c) of Scenarios 1 and 4, where
$X_{I}$ exists, and the PS model is linear. However, 
\mtt{UNI} has the advantage that when $X_I$ are
not useful for estimation, it does not use those variables for the
AIPW estimator. In the setting (b), where the PS model is misspecified, most of $X_I$ are dropped from $\widehat{\mathcal{U}}$, and \mtt{UNI} keeps efficiency as much as \texttt{OUT}. \texttt{\textsc{INT}}
is more variable under Setting (c) in Scenarios 1 and 3 since the $X_{P}$
are not considered. This phenomenon is consistent with the findings
in the previous research that including $X_{I}$ inflates standard
errors while including $X_{P}$ reduces standard errors \citep{shortreed2017outcome,tang2020outcome,rotnitzky2019efficient}.

In Setting (a)-(c), \texttt{O-UNI} and \texttt{UNI} are doubly robust
in the sense that it is unbiased, provided that either the OM model or the
PS model is correctly specified. Contrarily, \texttt{\textsc{INT}}
and \mtt{OUT} show large biases compared to \texttt{O-INT} and \texttt{O-OUT} in Setting (c). Theoretically, \texttt{INT} and \texttt{OUT} should be unbiased as in \texttt{O-INT} and \texttt{O-OUT} since the PS model is correct. However, \mtt{INT} and \mtt{OUT} are no longer unbiased because they do not consider $X_{I}$  necessary to maintain the double robustness of the AIPW estimator. These results show why we need to consider not only $X_C$ and $X_P$ but $X_I$ as well for the estimation when the OM model may be misspecified.
When both models are unknown, as in Setting (d), there is not much difference among \texttt{\textsc{UNI, INT}}, and \texttt{\textsc{OUT}} in terms of bias and efficiency because
most of the variables are not selected in the models.

Table \ref{tab: cr} displays the coverage rates for all scenarios. The results show that our coverage rates are close to the nominal coverage if either the OM or PS model is correctly specified. In contrast, the coverage rates for other approaches fail to reach the nominal
coverage rate if the OM model is misspecified.

\begin{table*}
\begin{center}
\caption{Simulation results for coverage rates}
\label{tab: cr} %
\begin{tabular*}{\textwidth}{@{\extracolsep\fill}lrrrrrr@{\extracolsep\fill}}
\toprule 
 & \mtt{O-UNI} & \mtt{O-INT} & \mtt{O-OUT} & \mtt{UNI} & \mtt{INT} & \mtt{OUT}\tabularnewline
\midrule
\multicolumn{7}{c}{Scenario 1} \tabularnewline \midrule
(a) OM I and PSM I & 0.939 & 0.944 & 0.942 & 0.938 & 0.946 & 0.942\tabularnewline
(b) OM I and PSM II & 0.954 & 0.945 & 0.952 & 0.953 & 0.960 & 0.952\tabularnewline
(c) OM II and PSM I & 0.950 & 0.950 & 0.945 & 0.950 & 0.929 & 0.873\tabularnewline
(d) OM II and PSM II & 0.791 & 0.918 & 0.788 & 0.788 & 0.922 & 0.786\tabularnewline
\midrule
\multicolumn{7}{c}{Scenario 2} \tabularnewline \midrule
(a) OM I and PSM I & 0.955 & 0.955 & 0.955 & 0.955 & 0.955 & 0.955\tabularnewline
(b) OM I and PSM II & 0.947 & 0.947 & 0.947 & 0.947 & 0.967 & 0.947\tabularnewline
(c) OM II and PSM I & 0.949 & 0.949 & 0.949 & 0.948 & 0.000 & 0.000\tabularnewline
(d) OM II and PSM II & 0.042 & 0.042 & 0.042 & 0.041 & 0.039 & 0.039\tabularnewline
\midrule 
\multicolumn{7}{c}{Scenario 3} \tabularnewline \midrule
(a) OM I and PSM I & 0.947 & 0.951 & 0.947 & 0.946 & 0.953 & 0.947\tabularnewline
(b) OM I and PSM II & 0.950 & 0.954 & 0.950 & 0.949 & 0.962 & 0.950\tabularnewline
(c) OM II and PSM I & 0.948 & 0.953 & 0.948 & 0.949 & 0.916 & 0.805\tabularnewline
(d) OM II and PSM II & 0.004 & 0.393 & 0.004 & 0.004 & 0.391 & 0.004\tabularnewline
\midrule
\multicolumn{7}{c}{Scenario 4} \tabularnewline \midrule
(a) OM I and PSM I & 0.941 & 0.945 & 0.945 & 0.941 & 0.945 & 0.945\tabularnewline
(b) OM I and PSM II & 0.945 & 0.945 & 0.945 & 0.945 & 0.962 & 0.945\tabularnewline
(c) OM II and PSM I & 0.940 & 0.952 & 0.952 & 0.941 & 0.008 & 0.008\tabularnewline
(d) OM II and PSM II & 0.622 & 0.621 & 0.621 & 0.627 & 0.623 & 0.623\tabularnewline
\bottomrule 
\end{tabular*}
\\
\end{center}
\emph{Note}: $X_I$ exists in Scenarios 1 and 4, $X_C$ exists in all Scenarios, and $X_P$ exists in Scenarios 1 and 3. Under OM I (II), the OM model is correctly specified (misspecified), and under PSM I (II), the PS model is correctly specified (misspecified).
\end{table*}

\section{Application}\label{6 application}


Low birth weight infants undergo severe health and developmental difficulties, which incurs enormous societal costs. Thus, considerable attention has been focused on finding the causal determinant of an infant's birth weight. Maternal smoking is a significant risk factor for low
birth weight infants \citep{kramer1987determinants,vogler2002differential}. Many studies were carried out to determine the relationship between maternal smoking during pregnancy and low birth weight infants. \citet{almond2005costs} implement a program evaluation approach. \citet{lee2017doubly} obtain a uniformly valid confidence band to show how smoking changes across different age groups of mothers.

The data is available on the STATA website\footnote{http://www.stata-press.com/data/r13/cattaneo2.dta}. The sample size for the data is 4262. The outcome of interest $Y$ is infant birth weight measured in grams. The treatment variable $A$ is
a binary variable equal to $1$ if the mother smokes and $0$ otherwise. We are interested in getting the ACE of maternal smoking during pregnancy on infant birth weight using the proposed method. We consider 17 covariates for analysis. The included covariates are an indicator of being married (\texttt{mmarried}), an indicator of Hispanic (\texttt{mhisp, fhisp}),
an indicator of foreign (\texttt{foreign}), an indicator of alcohol consumed during pregnancy (\texttt{alcohol}), an indicator of newborns died in previous births (\texttt{deadkids}), age (\texttt{mage, fage}), education attainment (\texttt{medu, fedu}), the number of prenatal
care visits (\texttt{nprenatal}), months since last birth (\texttt{monthslb}), the order of birth of the infant (\texttt{order}), race (\texttt{mrace, frace}), trimester of first prenatal care visit (\texttt{prenatal}), and the month of birth (\texttt{birthmonth}). Additionally, we add quadratic terms of the five continuous variables and 26 interaction
terms significant in either the PS model or the OM model. Therefore, the total number of covariates is 48.

\begin{figure}
\centering
\includegraphics[width=0.7\textwidth]{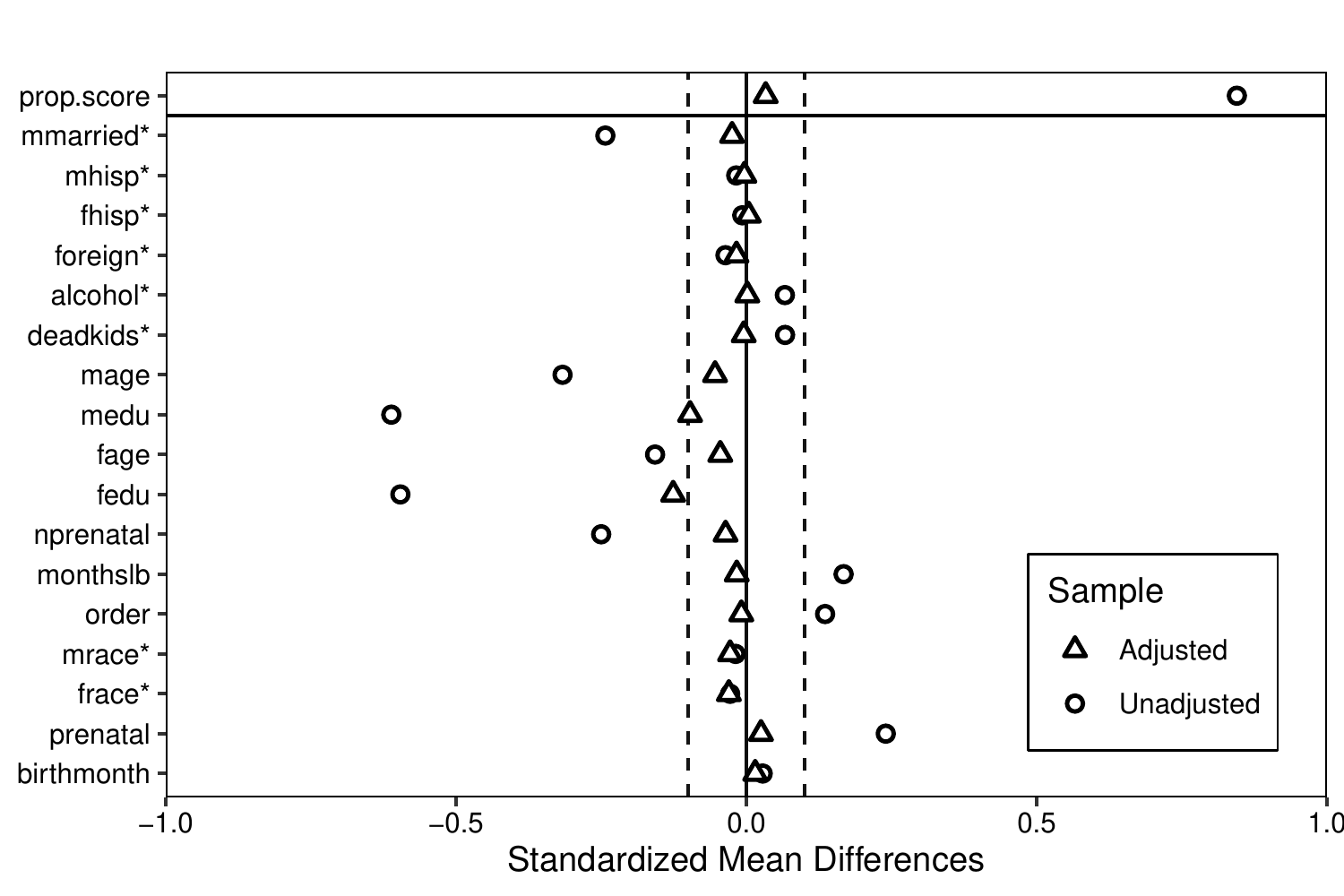}
\caption{Balance check: standardized mean difference for covariates of birth weight data. The dashed vertical lines are drawn at 0.1 SMD. }
\label{fig: pa love plot}
\end{figure}

Figure \ref{fig: pa love plot} displays the standardized mean difference for the covariates without an asterisk and the raw difference in means for the covariate with an asterisk. Note that the distribution of covariates is not balanced, which indicates the simple difference between the two treatment groups can introduce bias for the ACE. To estimate the ACE with our estimator, we assume the PS model to be a logistic regression model and the OM model to be a linear regression model. We estimate the standard errors using the asymptotic variance of Eq. (\ref{eq: TE of IF}).

Table \ref{tab: selection result pa} summarizes the selection results.
There are 15 instrumental variables, 21 confounding variables, and
four instrumental variables, which is similar to Scenario 1 in Section
\ref{5 simulation}. 

\begin{table}
\small
\centering
\caption{Selection result for birth weight data}
\label{tab: selection result pa}
\begin{tabular}{lllll}
\toprule
 & \multicolumn{4}{c}{Selected variables}  \\ \midrule
$X_I$ & \multicolumn{4}{l}{\begin{tabular}[c]{@{}l@{}}\texttt{mhisp,medu,fage,frace,birthmonth,mage,fedu,  }\\
\texttt{nprenatal:monthslb, nprenatal:order,alcohol:medu,}\\
\texttt{mmarried:foreign,mmarried:mage,foreign:nprenatal,}\\
\texttt{alcohol:fedu, medu:fedu}\\
\end{tabular}} \\ \midrule
$X_C$  & \multicolumn{4}{l}{\begin{tabular}[c]{@{}l@{}}\texttt{mmarried,mage,fedu,nprenatal,order,mrace,prenatal,}\\ \texttt{age,nprenatal,mmarried:mrace,alcohol:mage,}\\ \texttt{alcohol:nprenatal,deadkids:medu,monthslb:prenatal,}\\ \texttt{fhisp:order,foreign:mage,foreign:mrace, }\\ \texttt{mhisp:order,mmarried:fhisp,deadkids:prenatal,}\\
\texttt{frace:birthmonth}\end{tabular}}  \\ \midrule
$X_P$    & \multicolumn{4}{l}{\begin{tabular}[c]{@{}l@{}}\texttt{fhisp,foreign,deadkids:order,deadkids:birthmonth}\end{tabular}} \\ \midrule
             & \multicolumn{4}{c}{$p=48$, $|U|=40$, $|I|=21$, $|M_\beta|=25$}  \\ \bottomrule
\end{tabular}
\end{table}

Table \ref{tab:PA_result} displays the point
estimates, the standard errors, and the 95$\%$ Wald confidence intervals.
The result shows a similar pattern to Setting (c) in Scenario 1. \mtt{UNI}
has a larger standard error than \mtt{INT} and \mtt{OUT}. Also,
the estimate of \mtt{UNI} is different from \mtt{INT} and \mtt{OUT}.
As seen by the simulation in Section \ref{5 simulation}, \mtt{INT}
and \mtt{OUT} may be biased due to the use of a wrong set, while
the proposed method may correct the bias by its doubly robust property.
With the proposed estimator, maternal smoking reduces birth weight by 218.67g on average, which is a smaller decrease than those with \mtt{INT} and \mtt{OUT}. All $95\%$ confidence intervals do not include 0, which means it is significant at the 0.05 level that maternal smoking has a negative effect on birth weight.

\begin{table}
\centering
\caption{Point estimate, standard error, and 95\% Wald confidence interval for birth weight data}
\label{tab:PA_result}  %
\begin{tabular}{lllc}
\toprule
 & Est & SE & CI\tabularnewline
\midrule
\mtt{UNI} & -218.67 & 50.54 & (-317.73, -119.62)\tabularnewline
\mtt{INT} & -229.38 & 28.19 & (-284.62, -174.13)\tabularnewline
\mtt{OUT} & -229.94 & 27.31 & (-283.46, -176.43)\tabularnewline
\bottomrule 
\end{tabular}
\end{table}

\section{Concluding remarks}\label{7 sec:Concluding-remarks}
We establish the two-stage procedure to estimate the ACE with variable selection and the AIPW estimator. We compare the robustness of the AIPW estimator coupled with the union, intersection, and outcome predictor strategies using extensive simulation. Our method
is most robust, remaining consistent if either the OM model or the PS model is correctly specified. Other methods fail to be doubly robust under the misspecification of the OM model. When the instrumental variables are selected for estimation, our procedure may be more variable
than other approaches. However, the inefficiency is offset by including precision variables. Thus, when there are precision variables, the AIPW estimator based on the proposed variable selection strategy is less variable than that based on the intersection strategy. 
Our simulation results also imply that all strategies are badly biased in the case when the OM model and the PS model are misspecified. Thus, we still need a correct specification of either the PS or OM model for consistent estimation of our procedure. Although we employ the SCAD for penalization, our method is flexible in the sense that other penalization methods, such as LASSO or Minimax concave penalty, can be used to select variables at the first stage. In particular, when there is a high correlation among variables, introducing the $L_2$-penalty may improve the performance of Steps 1 and 2. Elastic net proposed by \citep{zou2005regularization} outperforms the LASSO, and the SCAD-$L_2$ performs better in terms of minimizing prediction error and maintaining variable selection precision than the SCAD \citep{zeng2014group}.

There are several directions for future work: (i) we will extend the results to the causal analysis of longitudinal observational studies \citep{yang2018semiparametric} and survival outcomes \citep{yang2020semiparametric}; and (ii) we will develop variable selection procedures when confounders are subject to missingness \citep{yang2019causal}, which is common-place in practice.

\section*{Acknowledgements}
Yang is partially supported by the NIA grant 1R01AG066883 and NIEHS grant 1R01ES031651.
\bibliography{aipw_review.bib}
\bibliographystyle{apalike}

%
%

\end{document}